\documentclass{aa} 
\usepackage{graphicx}
\usepackage{natbib}
\usepackage{longtable,lscape}
\usepackage{rotating}
\usepackage{fmtcount}
\usepackage{amsmath}
\usepackage{adjustbox}
\usepackage{txfonts}
\usepackage{hyperref}

\newcommand{\teff}{\mbox{$T_{\rm eff}$ }}
\usepackage{color}

\begin{document}

\title{New \teff and [Fe/H] spectroscopic calibration for FGK
dwarfs and GK giants} 
\subtitle{}
   \author{G. D. C. Teixeira \inst{1,2}, S. G. Sousa\inst{1}, M. Tsantaki \inst{1,3}, M. J. P. F. G. Monteiro \inst{1,2},  N. C. Santos \inst{1,2} \and G. Israelian\inst{4,5}
                }
   \offprints{G. D. C. Teixeira; email:gteixeira@astro.up.pt}
   \institute{ Instituto de Astrof\'isica e Ci\^encias do Espa\c{c}o, Universidade do Porto, CAUP, Rua das Estrelas, 4150-762 Porto, Portugal
         \and
         Departamento de F\'isica e Astronomia, Faculdade de Ci\^encias, Universidade do Porto, Rua do Campo Alegre, 4169-007 Porto, Portugal
         \and Instituto de Radioastronom\'ia y Astrof\'isica, IRyA, UNAM, Campus Morelia, A.P. 3-72, C.P. J8089 Michoac\'an, Mexico
      \and Instituto de Astrof\'{i}sica de Canarias, 38200 La Laguna, Tenerife, Spain 
      \and Departamento de Astrof\'{i}sica, Universidad de La Laguna, 38206 La Laguna, Tenerife, Spain}
 \authorrunning{Teixeira et al.}
   \titlerunning{New spectroscopic calibration for FGK dwarfs and GK giants}
\date{\today}
 
  \abstract
 {The ever-growing number of large spectroscopic survey programs has increased the importance of fast and 
 reliable methods with which to determine precise stellar parameters. Some of these methods are highly dependent on correct spectroscopic
 calibrations.}
{The goal of this work is to obtain a new spectroscopic calibration
for a fast estimate of \teff and [Fe/H]
for a wide range of stellar spectral types.}
{We used spectra from a joint sample of 708 stars, compiled from 451 FGK dwarfs and 257 GK-giant
stars. We used homogeneously determined spectroscopic stellar parameters to derive
temperature calibrations using a set of selected EW line-ratios, and [Fe/H] calibrations
using a set of selected \ion{Fe}{i} lines.}
{We have derived 322 EW line-ratios and 100 \ion{Fe}{i} lines that can be used 
to compute \teff and [Fe/H], respectively. We show that these calibrations are
effective for FGK dwarfs and GK-giant stars in the following ranges:
$4500 \ K < \teff < 6500 \ K$ , $2.5 < \log g < 4.9 \ dex$, and $-0.8 < [Fe/H] < 0.5 \ dex$.
The new calibration has a standard deviation of 74 K for \teff and $0.07 \ dex $ for [Fe/H].
We use four independent samples of stars to test and verify the 
new calibration, a sample of 56 giant stars, a sample composed of Gaia FGK benchmark
stars, a sample of 36 GK-giant stars of the DR1 of the Gaia-ESO survey, and a sample of 582 FGK-dwarf stars.
We also provide a new computer code, GeTCal, 
for automatically producing 
new calibration files based on any new sample of stars.}
{}

   \maketitle
%

\section{Introduction}
        
        The derivation of accurate and precise stellar parameters
        has become a fundamental aspect of astrophysical
        studies. Parameters such as stellar mass, radius, and age
        are tremendously
        important, but very difficult
        to measure 
        directly.

        Stellar parameters are important for a diverse range of astrophysical
        studies: from characterization of planet-host stars to Galactic
        population studies
        (\citet{edv93,fisch05,casa10,hub12,mortier13,bens14,chap14}, to name a few).
        The determination of mass, radius, and age requires a knowledge of
        stellar atmospheric parameters, such as
        effective temperature (\teff), surface gravity ($\log g$),
        metallicity ([Fe/H]), and microturbulence($v_{mic}$), which can be obtained mainly by
        spectroscopic or photometric methods (e.g. 
        \citet{sousa451stars,casa10,maria13}).

        Large-survey observational programs such as GES \citep{ges}, GALAH \citep{gala}, and RAVE \citep{rave}  are quickly becoming
        the new paradigm of stellar observations and characterization. 
        Given the volume of data involved, the success of these programs relies on the existence of methods
        that rapidly ascertain stellar parameters for a diverse range of stellar spectral types. 
        Such methods cover a broad range of evolutionary
        stages from
        main-sequence dwarf stars to post-main-sequence giants and all the intermediate
        stage stars.
        Since it is difficult to effectively cover such a diversity of
        evolutionary stages in a uniform way, broad-range methods are rare.
        These stars can be studied by methods like the line-strength ratios
        to obtain \teff \citep{gray96, graybook, kov03}, and equivalent widths (EWs) of \ion{Fe}{i}
        to derive [Fe/H] \citep{sousaTMCALC}.
        One of the most important steps in building these methods is to obtain an
        empirical calibration.

        One of the spectroscopic methods that can be used to determine 
        spectral parameters is the ARES+MOOG method.
        This method computes stellar atmospheric
        parameters based on the use of two codes: ARES and MOOG.
        The ARES code is an automated tool for obtaining the EWs of a spectrum based on 
        an initial line list, and also weights parameters like the signal-to-noise ratio (S/N) \citep{sousaARES,ares2015}.
        MOOG is able to perform a variety of spectral line analyses 
        and synthesis computations under local thermodynamic equilibrium (LTE) 
        conditions \citep{sneden73moog}. 
        In the ARES+MOOG method, MOOG is used to measure individual line abundances
        and is combined with a minimization algorithm based on the simplex method to derive the parameters of the stellar atmosphere that best 
        fits the measured EWs.
        For a more detailed description of this method see
        \citet{sousaschool}.

         \citet{sousaTMCALC} presented an automated tool, TMCalc, with which
\teff and [Fe/H] can be obtained extremely fast using measurements of
        EWs of spectral lines. It is based on EW line-ratios 
        and on the EWs of \ion{Fe}{i} lines.
        The accuracy and precision of the results produced by TMCalc 
        are mainly limited by the \teff and [Fe/H] calibrations, the EW measurements, and
        the (S/N) of the stellar spectra \citep{sousaTMCALC}.
        The goal of our work is to produce a new calibration that is compatible with
        a broader regime of stellar evolutionary stages.
        
        The main difference between using TMCalc and the ARES+MOOG method is 
        that with TMCalc we obtain \teff and [Fe/H] by a simple application of 
        EW line ratios, while in ARES+MOOG a minimization procedure is performed 
         to obtain the stellar atmospheric parameters. TMCalc is  therefore
        less accurate, but it is computationally more efficient.
        
        The paper is organized as follows: in Sect. \ref{ch:samples} we present the 
        stellar samples we used and 
        the updated parameters. In Sect. \ref{ch:calibration} we
        explain
        the different steps and 
        assumptions for \teff and [Fe/H] calibrations,
        and we present an automatic calibration tool: GeTCal. 
        In Sect. 
        \ref{ch:results} we compare the new calibration with the pre-existing one,
        show the parameters obtained for two independent samples, and make some considerations
        on the effect that a poor \teff determination may have on [Fe/H] calculations.
        Finally, in Sect. \ref{ch:conclusions} we summarize our results.

\section{Stellar samples}\label{ch:samples}
        
        Our work was performed using six distinct stellar samples:
        two samples were used for calibration and 
        four other samples were used for 
        independent testing of the new calibration. A summary of each sample can be
        found in Table \ref{tab:sample_comparison}, where we provide the number of stars,
        (S/N), spectral classification, and intervals in \teff, $\log g$,
        $v_{mic}$ , and [Fe/H] for each sample. 
        
        All samples parameters have been consistently determined
        homogeneously with ARES+MOOG. 
        
        We measured the EWs for all stars in the six samples
        with the initial line list of \citet{sousa10}        to obtain homogeneous and comprehensive measurements and 
        taking into consideration the different (S/N) 
        of each spectrum.
        
        ARES performs a local normalization before measuring the 
        EWs. The errors on the measured EWs depend on the different (S/N), and 
        line-blends are taken into account since ARES performs multi-line 
        fits \citep{sousaARES}. Although the new version of ARES reports 
        errors on the EWs \citep{ares2015}, this work made use of the previous version.
        
        \subsection{Calibration samples}
        
        To obtain the new calibration, we used two samples:
        
        \begin{itemize}
         \item The sample of well-studied 451 FGK-dwarf 
        stars described in \citet{sousa451stars} that was revised in
        \citet{maria13}, hereafter the So08 sample.

         \item The sample of 257 giant stars from \citet{alves14}, hereafter the Al15 sample.

        \end{itemize}
                 
        The So08 sample is composed of high-quality spectra of 
        451 FGK-dwarf stars with well-determined parameters \citep{sousa451stars,maria13}.
        These stars have been analysed using HARPS high-resolution spectral data with
        a resolution $R \sim 110000$ and an (S/N) ranging from $\sim 70$ to
        $\sim 2000$. 
          The parameters for this sample were revised in \citet{maria13}
to address an overestimation of
        \teff for stars in the low-temperature regime, $\teff < 5200 \ K$.
        In the remainder of this work, when we refer to the So08
sample, we
        refer to the corrected sample. 
        
        The Al15 sample is composed of high-resolution spectra obtained with 
        the UVES spectrograph 
        for 257 GK-giant stars. The spectra have resolutions of $\sim 110000$ with an (S/N)
        of $\sim$ 150. The parameters for these stars have been determined by \citet{alves14}
        using the same standard analysis of ARES+MOOG as in this work.
        
        A joint sample, composed of the So08 and Al15 samples, was used as the calibration
        sample for our study, hereafter the joint sample.

        \begin{table*}
          \centering
          \caption{ Summary of the stellar samples.}
         \label{tab:sample_comparison}
         \begin{tabular}{lcccccccc}
         \hline
         \hline
         
         Sample & stars & stars\_used &(S/N) & Spec type & \teff  & $\log g$  & $v_{mic}$  & [Fe/H] \\ 
           &   &   & & & ($K$) & ($dex$) &  ($km s^{-1}$) & ($dex$)\\
         \hline
         So08    & $451$ & $451$   & $[70,2000]$    & FGK        & $[ 4400, 6431]$ & $[ 3.60, 4.82]$ & $[ 0, 2.1]$     & $[ -0.83, 0.36]$ \\ 
         Al15    & $257$ & $257$   & $\sim 150$     & GK         & $[ 4724, 5766]$ & $[ 2.37, 3.92]$ & $[ 1.08, 4.28]$ & $[ -0.75, 0.27]$ \\
         Sa09    & $56$  & $44$   & $\sim 200$     & GK         & $[ 4157, 6020]$ & $[ 1.15, 4.82]$ & $[ 0.86, 2.26]$ & $[ 0, 0.32]$     \\ 
         Gaia    & $34$  & $18$   & $[200,400]$    & FGK        & $[ 3472, 6635]$ & $[ 0.51, 4.67]$ & $[ 0.89, 1.92]$ & $[ -2.64, 0.35]$ \\
         GES     & $36$ &  $36$   & $\sim 100$    & GK         &$[ 4753, 5289]$ & $[ 2.50, 3.86]$&  $[ 1.17, 4.06]$  &  $[ -0.51, 0.25]$  \\
         So11    & $582$        &  $582$   & $[100,200]$   & FGK        & $[ 4487, 7212]$ &  $[ 3.61, 4.96]$  &  $[ 0, 2.87]$  &  $[ -1.14, 0.55]$  \\         
         \hline
         \end{tabular}
         \end{table*}

        \subsection{Validation samples}
        
        To test the new calibration, we used four distinct and independent samples:
        
        \begin{itemize}
        
         \item The sample of 44 giant stars from \citet{santos09}, hereafter the Sa09 sample.

         \item A sample of 18 benchmark stars of the Gaia survey, hereafter the Gaia sample \citep{jofre14, heit15}.
         
         \item A subsample of 36 GK-giant stars from the Gaia-ESO survey DR1 \footnote{\url{https://www.gaia-eso.eu/sites/default/files/file_attach/ESO-DR1-release-description.pdf}}  with 
         $\log g <3.9$, hereafter the GES sample.
         
         \item A sample of 582 FGK-dwarf stars from \citet{sousa11} with 
         with well-determined parameters, hereafter the So11
sample.

        \end{itemize}
        
        {The histograms with the distribution of \teff, $\log g$, and [Fe/H] of the validation samples is shown in Fig. \ref{fig:validation_sample}.
        
        The Sa09 sample of 56 giant stars has been observed using the UVES spectrograph,
        with a spectral resolution of between $R\sim 50000$ and $\sim 100000$
        and an (S/N) of $\sim 200$.
        The parameters used for the Sa09 sample were rederived in this work using 
        the ARES+MOOG method. Using the applicability criteria, which we        discuss in detail in Sect. 4.1, this sample was reduced to 44 stars within 
        our applicability limits.
        
        The Gaia sample is composed of the 34 FGK benchmark stars
        with well-determined values \citep{jofre14}. Of these, we used a 
        sub-sample of 28 stars since the remaining six stars were M dwarfs and
        therefore unsuitable for EW automatic measurements. A final selection 
        was applied  to have only stars that fulfilled our applicability
        criteria, leading to 18 benchmark stars.
        The spectra have 
        a high resolution (from HARPS and NARVAL spectrographs) and high (S/N), 
        with values ranging from $\sim 200$ to $\sim 400$. The parameters of
        this sample were obtained by combining the EW methods of several work groups 
        in the Gaia-ESO survey \citep{jofre14, heit15}.
        
        The GES subsample is composed of 36 GK-giant stars. The data used were taken from 
        the Gaia-ESO survey DR1 spectra. 
        The parameters for this sample have been derived using ARES+MOOG within 
        the Porto/CAUP node in GES. 
        The stars in these sample were observed for GES with the UVES spectrograph in the 580 nm setup.
        The (S/N) of the spectra was $\sim 100$.

        The So11 sample is composed of 582 FGK-dwarf stars. The data used were taken from 
        the So11 spectra. The (S/N) of the spectra ranged from $\sim 100$ to $\sim 200$. The EWs and parameters of these stars 
        were presented in \citet{sousa11} using the ARES+MOOG method.
        
        The Gaia, GES, So11, and Sa09 sample were used as independent samples
        to test our new calibration (see Sect. \ref{subs:indsamp} ).

        \begin{figure}
        
         \includegraphics[width=\columnwidth]{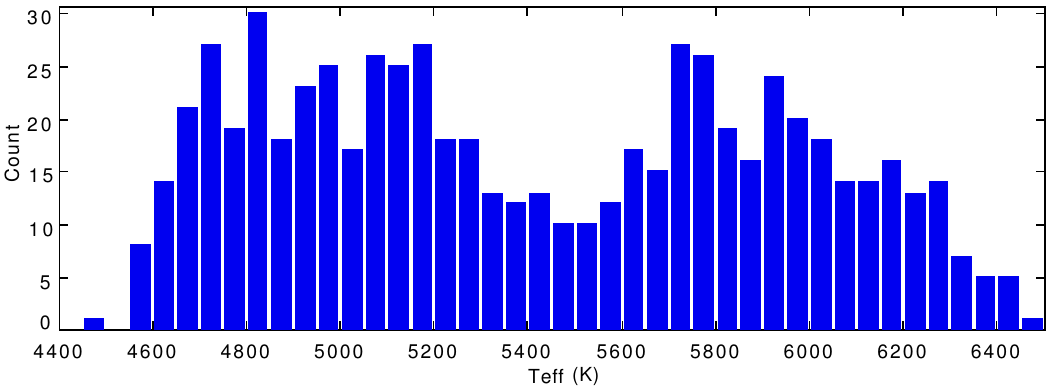}
         \includegraphics[width=\columnwidth]{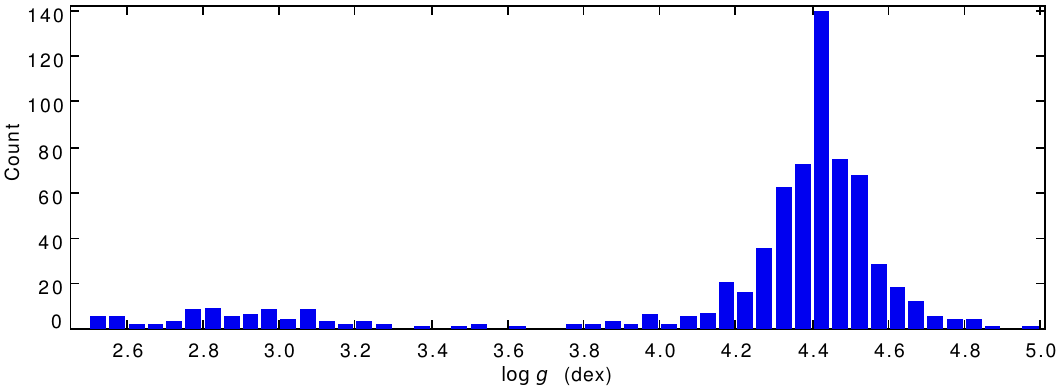}
         \includegraphics[width=\columnwidth]{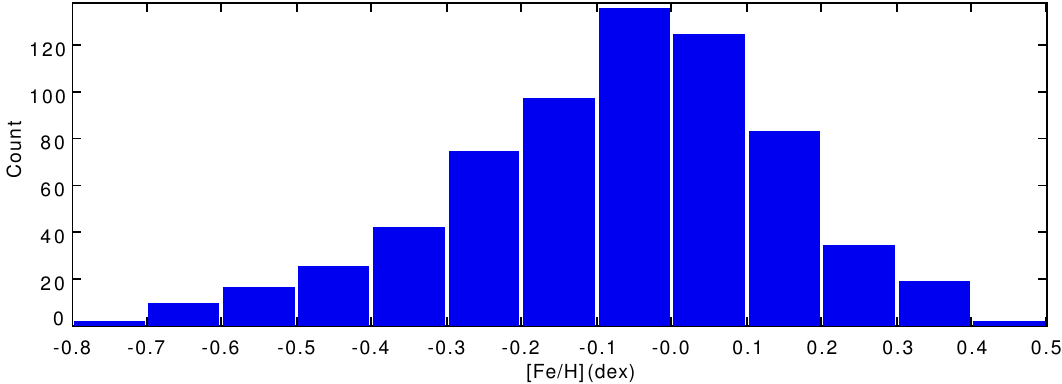}
         \caption{Histograms showing the parameter distribution of the validation sample.
         From top to bottom we show the \teff, $\log g$, and [Fe/H] determined with spectroscopic methods.}
        \label{fig:validation_sample}
        \end{figure}    
        
\section{Calibration}\label{ch:calibration}
        
        The main objective of this work is to build new \teff and [Fe/H] spectroscopic
        calibrations for both FGK dwarfs and GK giants.
        This new calibration will be an improvement over the 
        calibration presented in \citet{sousaTMCALC}, hereafter the So12 calibration.

        Here we present the \teff calibration, 
        the procedure of the [Fe/H] calibration and, finally, we describe an
        automatic code for producing these calibrations.

        \subsection{\teff calibration}\label{ch:teffcal}

        To perform the \teff calibration, we used the relations between EW line-ratios, closely
        following \citet{sousa10}.
        The basis of this technique is that metal lines have 
        different sensitivities to \teff and can be used in a similar 
        way as in spectral type determinations. EW line-ratios are more precise
        than the use of EWs of individual lines 
        \citep{gray96,graybook,kov03,sousa10}.
        
        To fully exploit the advantages of EW line-ratios, some 
        considerations were 
        taken into account: 
        
        \begin{itemize}
         \item The difference in excitation potential of the lines should be greater than 3 eV, 
         ensuring that the 
         lines have a 
         different  
         sensitivity to
         \teff changes.
         
         \item The lines used for the ratios should not be too distant in wavelength ($\Delta \lambda < 70 \AA$) to 
         minimize errors in continuum determination.
         
        \end{itemize}

        After compiling ratios that fulfilled the conditions described above for 
        each star in the sample, we 
        proceeded to refine the chosen ratios using additional selection
        criteria. We discarded ratios between lines that differed
by more than 
        two orders of magnitude in EW, thus removing ratios affected by poor
        measurements that are due to either poor Gaussian
        fitting or poor continuum fitting.

        We also implemented a new selection cut-off based on the interquartile 
        range method, IQR. This is a 
        reliable method of statistical analysis that
        uses the statistical dispersion to effectively trim outliers \citep{upton96}. This has not been implemented in the previous works.
        The IQR 
        measures the distance between the first and third quartiles
        of a distribution, $Q1$ and $Q3,$ respectively, 
        and only considers values in the interval
        
        \begin{equation}
         [Q1-1.5 \times IQR,Q3+1.5 \times IQR]
        .\end{equation}
        
        The IQR method was used to remove the outliers in the distribution 
        of each EW line-ratio, R\_EW.
        
        After the initial outlier-removal procedures, two functions were fitted to  
        distributions of 
        \teff as a function of EW line-ratios: a linear function and
        a third-degree polynomial function. An additional outlier-removal procedure 
        was then used based on a typical 2-$\sigma$ cut. 
        Subsequently, we refitted the functions and obtained their 
        coefficients. 
        Figure \ref{fig:sigma-cut} shows the 2-$\sigma$ cut procedure for the ratio
        between line \ion{Si}{i} (6142.49 \AA) and \ion{Ti}{i} (6126.22 \AA). 
        This particular ratio was chosen as a consistency check to
        show how well our procedure compares with the one presented in 
        Fig. 2 of \citet{sousa10}.
        
        \begin{figure}
        
         \includegraphics[width=\columnwidth]{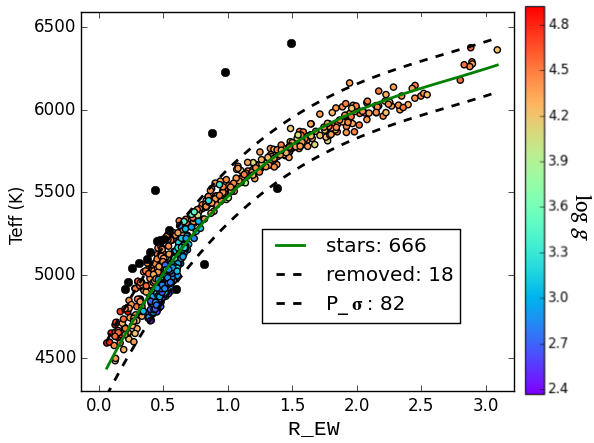}
         \caption{Application of the 2-$\sigma$ cut to the ratio  between the EW of  line \ion{Si}{i} 
         (6142.49 \AA) and of the EW of  \ion{Ti}{i} (6126.22 \AA).
         The green line represents the third-degree polynomial fit, the dashed blue line represents the 
         2-$\sigma$ interval, and the black dots are the stars removed by the procedure. The vertical axis
         represents the spectroscopic \teff and the horizontal axis represents the value of the EW ratios.
         The value of the final $\sigma$ of the fit, P\_$\sigma$, is $\sim 82 \ K$. The $\log g$ of each star is colour-coded. }
        \label{fig:sigma-cut}
        \end{figure}

        Following 
        \citet{sousa10}, we
        then applied these methods  
        to the inverse of the EW line-ratios, $1/R\_EW$, and to the
        logarithm of the EW line-ratios, $\log R\_EW$. 
        The values of the standard deviation of every
        fitted function mentioned above were compared to select the function 
        with the lowest standard
        deviation. 
        This fitting procedure is shown in Fig. \ref{fig:ratios}.
        The chosen function was the third-polynomial fit
        of the ratio R\_EW, with a $\sigma  \sim 86 \ K$.

        An additional validity check was implemented 
        by accepting only functions that fitted two-thirds
        of the calibration sample.
        This ensured that each EW line-ratio used
        was valid for a significant number of the stars 
        used in the calibration.

        \begin{figure}
        
         \includegraphics[width=\columnwidth]{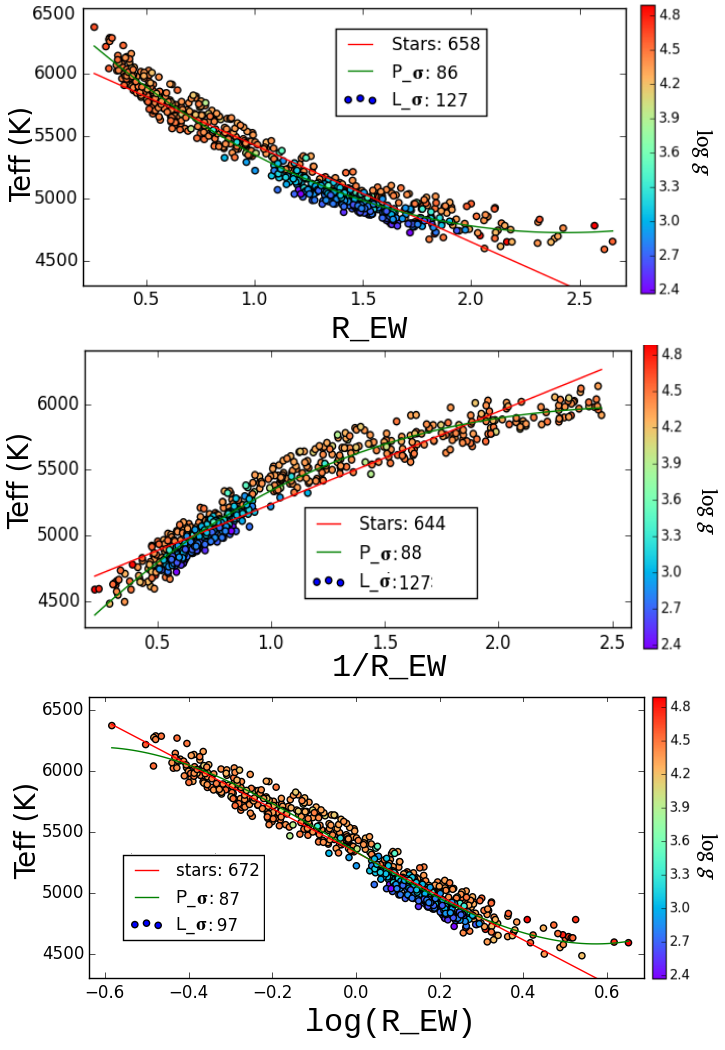}
         \caption{Fitting procedure for the \teff calibration using
         the ratio, R\_EW, between the EW of lines \ion{V}{i} (5670.85 \AA) 
         and \ion{Fe}{i} (5635.83 \AA). We also show the fit for the inverse of the ratio, $1/R\_EW$, and
         the logarithm of the ratio, $\log R\_EW$. The final number of stars
         used in the fit, the standard deviation of the linear fit, L\_$\sigma$, and of the 
         third-degree
         polynomial fit, P\_$\sigma$.  The $\log g$ of each star is colour-coded.}
        \label{fig:ratios}
        \end{figure}

        \subsection{Metallicity calibration}\label{ch:metalcal}
        
        We calibrated the metallicity using only iron absorption lines,
        since we used the iron abundance as a proxy for stellar metallicity \citep{sousaTMCALC}.
        We discarded any ionized iron lines from the list because
of their expected dependence
        on $\log g$.
        
        The dependence on microturbulence was minimized by considering
        only weak lines (EW $<$ 
        70 $m\AA$). 
        We also excluded lines with EW smaller 
        than 20 $m\AA$,
        which minimized errors in the EW measurements that are due to the increased difficulty
        in estimating the continuum. 
        
        For each line, the calibration was obtained from the following equation:
        
        \begin{equation}
        \label{eq:metal}
        \begin{split}
         EW = & C_0 + C_1 \times [Fe/H] + C_2 \times \teff + C_3 \times [Fe/H]^{2} + \\
         & C_4 \times \teff^{2} +C_5 \times [Fe/H] \times \teff
        \end{split}
        .\end{equation}
        
        This equation represents the simple dependence of line strength on effective temperature
        and iron abundance and was solved for [Fe/H] by inverting the 
        equation \citep{sousaTMCALC}. It is trivial to conclude that there will be a \teff 
        dependency
        for the [Fe/H] obtained from the inverted form of this equation.

        We computed the [Fe/H] from Eq. \ref{eq:metal} 
        for each line and 
        compared it with the 
        spectroscopic values.
        The outliers were removed by first applying the 
        IQR method,
        and a 2-$\sigma$ 
        cut was then applied. The outliers obtained with this
        method are usually due to poorly measured EWs, caused by strong blending effects
        or poor continuum determination.

        Two additional selection
        criteria were applied: 
        
        \begin{enumerate}
         \item The slope of the 
         comparison between the calibrated and spectroscopic
         [Fe/H] 
         has to be within 3\% of the identity line.
         
         \item Only standard deviations of individual line calibrations lower than 0.06 dex
         were considered.
        \end{enumerate}
        
        These conditions were empirically determined 
        to achieve a 
        balance between the largest number
        of lines possible and 
        the highest precision possible. 
        This balance ensures that we obtained statistical reliability while at the same time
        maintaining a high precision and accuracy in our calibration.
        
        \subsection{GeTCal}\label{ch:getcal}
        
        A useful by-product of our work in obtaining new calibrations was the 
        creation of a Python code: GeTCal. GeTCal is a pratical implementation of the 
        methods described in Sects. \ref{ch:teffcal} and \ref{ch:metalcal} and is capable of
        automatically producing \teff and [Fe/H] calibrations. It requires three input parameters:
        a line list, the stellar parameters (and errors) of a sample of stars, and the measured EWs
        for each star. 
        The calibrations can be
        used to compute the \teff and [Fe/H] of stars of similar spectral classes.
        It is capable of performing the \teff calibration, the [Fe/H] calibration, 
        or both simultaneously. 
        
        The GeTCal code is built in such a way as to produce calibration files
        that are compatible with the TMCalc code \citep{sousaTMCALC}.
        This code is freely distributed
        and available for use by the community
        \footnote{The GeTCal code and
        the \teff and [Fe/H] calibrations are available at
        \url{http://www.astro.up.pt/exoearths/tools.html}.}.
        
\section{Results}\label{ch:results}
        
        In this section, we present the new calibration and
        test its application with TMCalc. 
        We show consistency checks by comparing results obtained
        with the new calibration with those obtained 
        with the So12 calibration for the joint sample. 
        We present  
        and analyze
        the results obtained when applying 
        the new calibration to four independent stellar samples.
        Finally, we present and discuss 
        the effect that an erroneous \teff can have on
        [Fe/H] determination.
        
        \subsection{New calibration}\label{subs:newcal}
        
        Since our main goal is to obtain a new and more precise
        calibration with an increased range of applicability, accommodating 
        both FGK dwarfs and GK-giant stars, we used
        the joint sample for calibration procedures.
        
        Table \ref{tab:cal_comparison} shows a summary of both calibrations.
        A total of 322 EW line-ratios and 101 \ion{Fe}{i}
        are used in the new calibration, which is fewer than in the So12 calibration. 
        The reason is that we consider stars in different
        evolutionary stages and, therefore, there are lines and ratios that are poor \teff and [Fe/H] tracers for giants and dwarfs.
        
        We calculated the \teff using the So12 calibration for the joint sample.
        The top panel of Fig. \ref{fig:teff_orig_joint} compares
        the computed \teff with the spectroscopic \teff. 
        This calibration
        clearly does not fit either dwarfs in the low-temperature regime or
        giant stars properly, with a standard error of $112 \ K$ and a mean difference of $-90 \ K$ in
        \teff. This behaviour 
        was expected: the So12 calibration was created using only 
        the So08 dwarfs, without the  correction for the 
        low-temperature regime \citep{maria13}.

        \begin{figure}
        
         \includegraphics[width=\columnwidth]{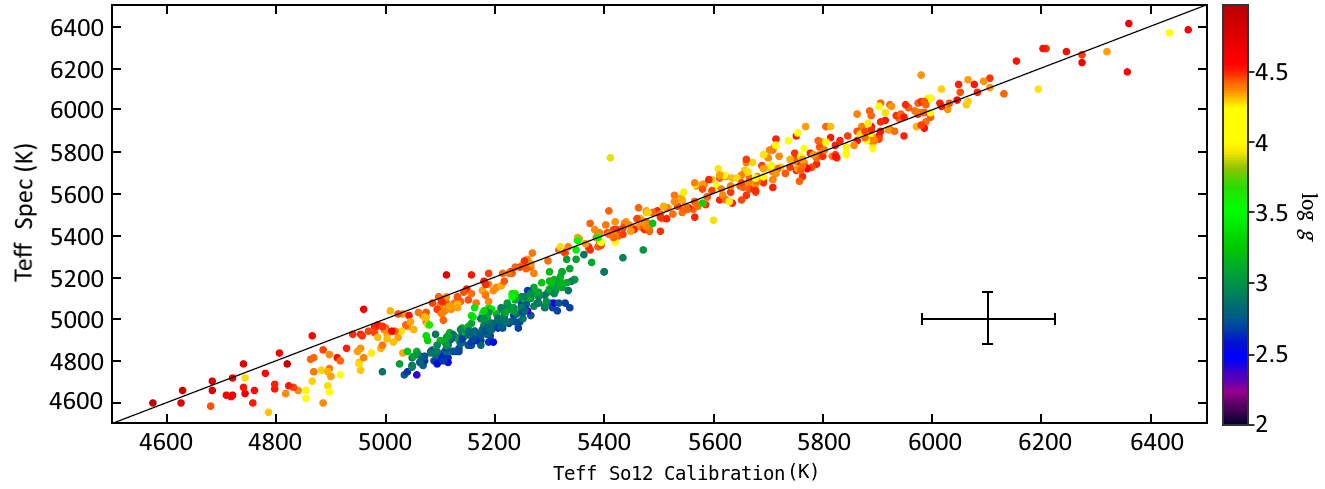}
         \includegraphics[width=\columnwidth]{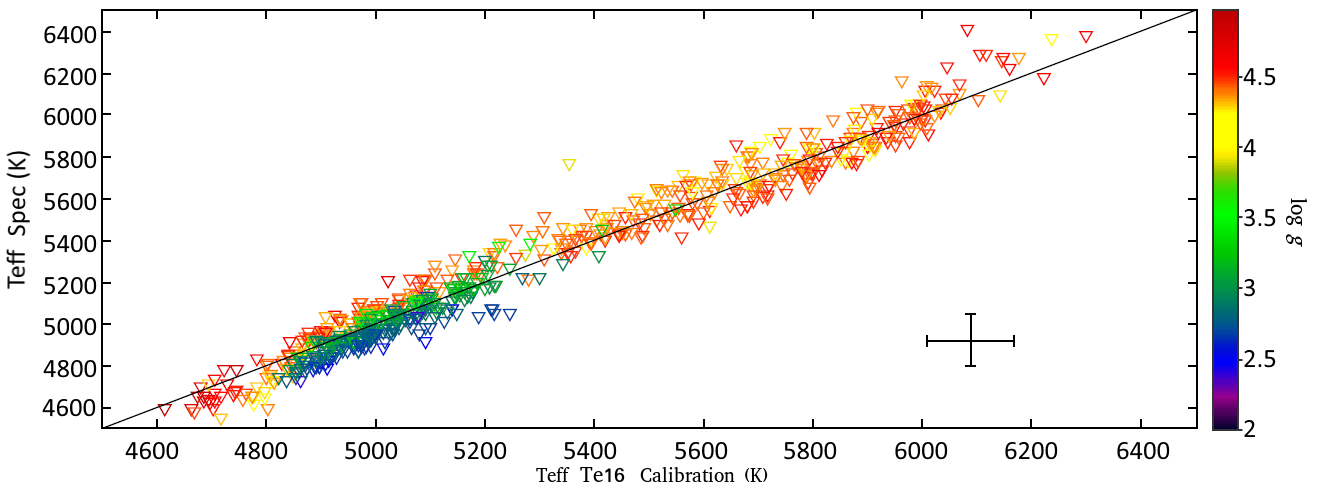}
         \includegraphics[width=\columnwidth]{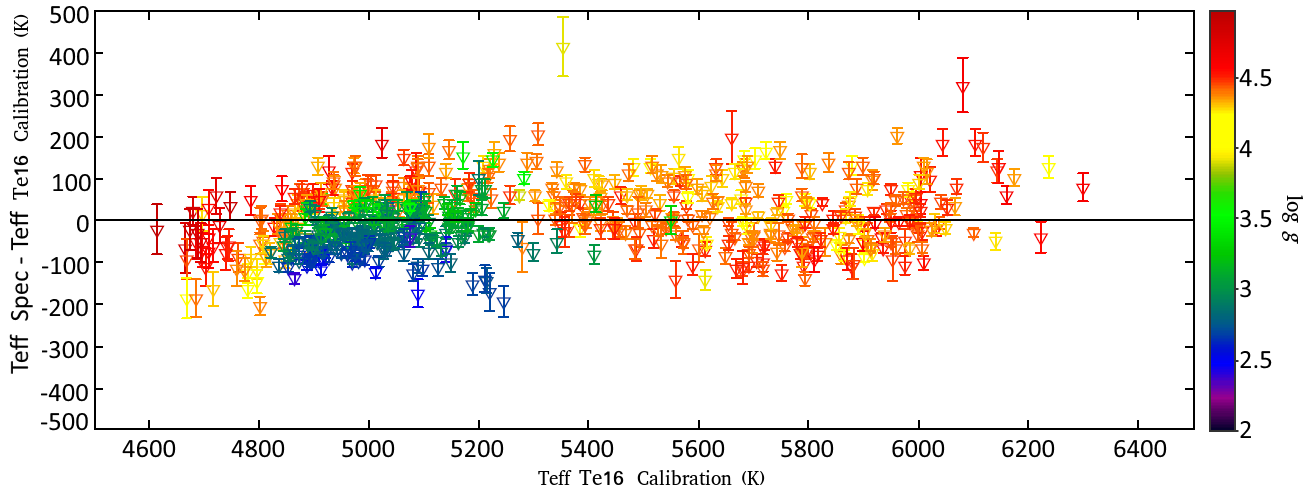}
         \caption{Comparison between the \teff computed in this work and the spectroscopic
         values with the So12 calibration (top panel) and the Te16 calibration (middle panel) 
         for the joint sample. \
         The line in the two plots represents the identity line, the standard deviation
         is plotted as the cross in both panels. \
         In the bottom panel we show the difference between the Te16 calibration and the spectroscopic
         \teff as a function of \teff, the error bars represent the errors in our computation. \
         The plots are colour-coded for $\log g$.}
        \label{fig:teff_orig_joint}
        \end{figure}
        
        The middle panel of Fig. \ref{fig:teff_orig_joint} shows the comparison of the 
        temperature determination using the newly computed calibration (hereafter
        the Te16 calibration). There is a clear improvement compared to the 
        So12 calibration, now with a standard deviation
        of $74 \ K$ and a mean difference of the computed values of $-1 \ K$.
        
        The results for the metallicity computation of the joint
sample are shown in
        Fig. \ref{fig:feh_orig_joint}.
        The So12 calibration forms a completely
        distinct locus for the GK-giant stars (top panel). On the other hand, 
        the Te16 calibration removes the
        different locus for the GK giants of the sample (middle panel). 
        We also plot 
        the difference between the spectroscopic and computed
        [Fe/H] as a function of [Fe/H] (bottom panel). Most 
        stars are well within $0.2$ dex from the zero value.
        
        \begin{figure}
        
         \includegraphics[width=\columnwidth]{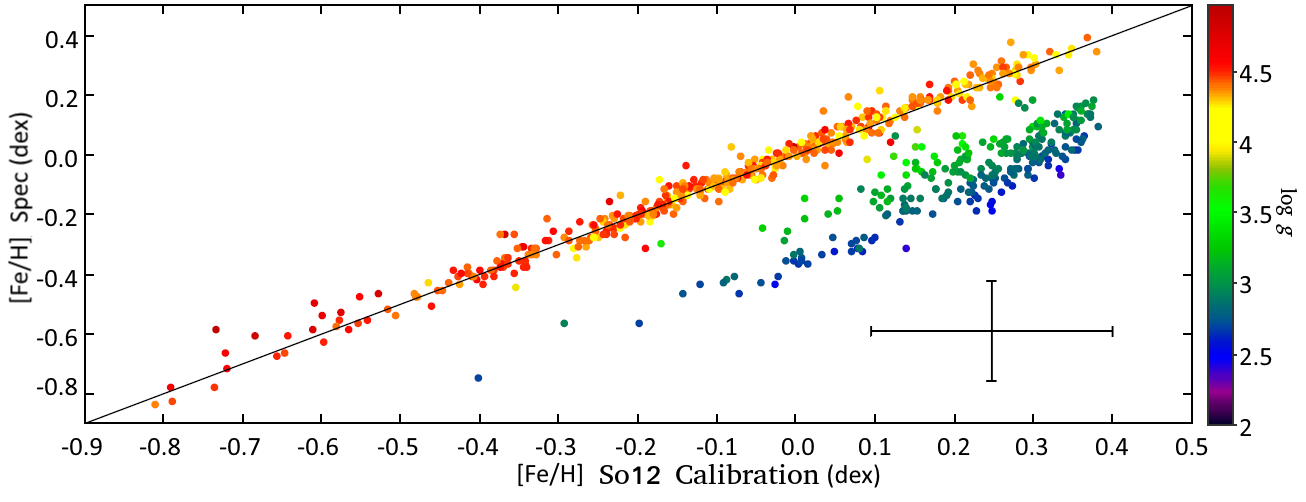}
         \includegraphics[width=\columnwidth]{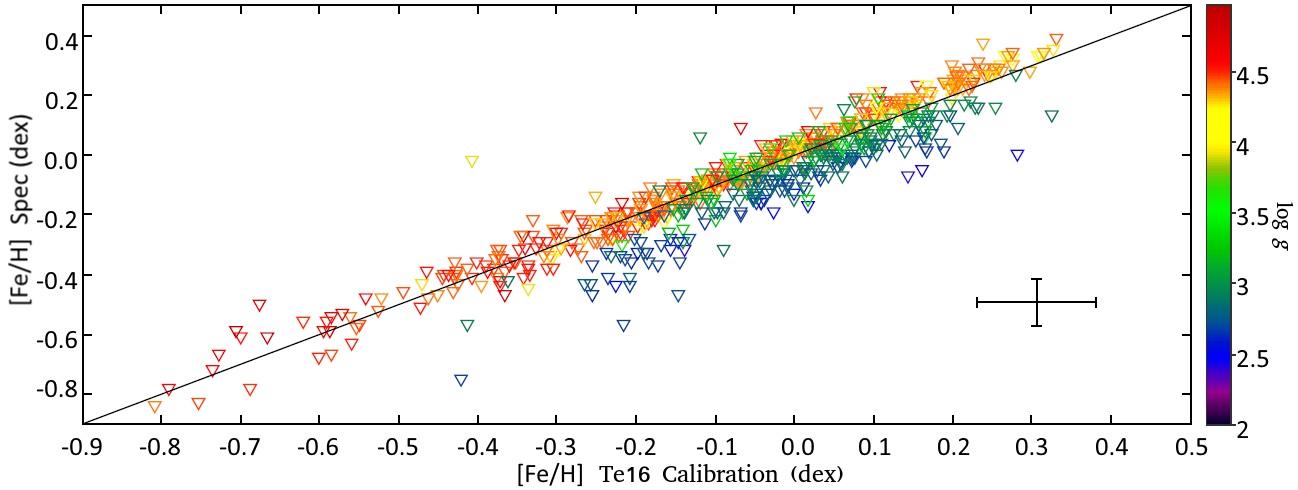}
         \includegraphics[width=\columnwidth]{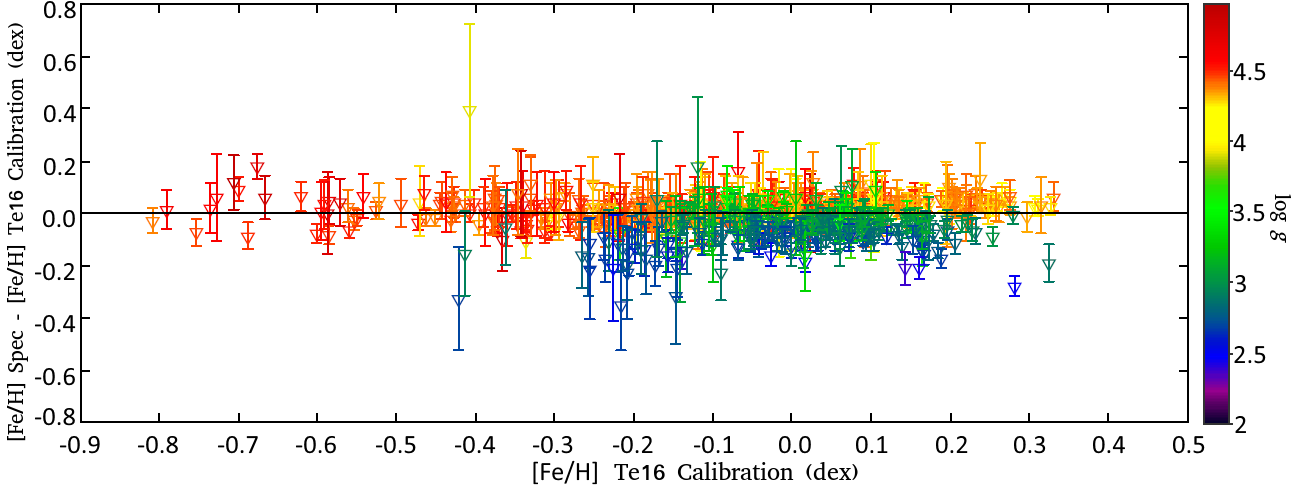}
         \caption{Comparison between the [Fe/H] computed in this work and the spectroscopic
         values for the So12 calibration (top panel) and the Te16 calibration (middle panel) for the joint sample. \
         The black line represents the identity line, the standard
         deviation is shown as the cross. The bottom panel shows the difference between [Fe/H] from spectroscopy 
         and the one in this work as a function of [Fe/H], the error bars are the errors resulting from this work.\
         The plots are colour-coded for $\log g$.}
        \label{fig:feh_orig_joint}
        \end{figure}

        Table \ref{tab:full_comparison} shows the comparison between the        two calibrations for the So08, Al15, and the
joint samples. 
        The mean difference between the spectroscopic and computed \teff improves from
        $-90$ to $-1 \ K$ when we apply the So12 calibration and the Te16 calibration for the joint sample, respectively.
        Likewise, there is an improvement in the mean difference in the values of [Fe/H],
        which is $-0.10 \ dex$ for the So12 calibration and $-0.02 \ dex$ for the Te16 calibration.
        The value of the computed \teff standard deviations changes from $112 \  K$ in the So12
        calibration to $74 \ K$ in the Te16 calibration, and the standard deviation 
        of [Fe/H] improves from $0.15 \ dex$ in the So12
        calibration to $0.07 \ dex$ of the Te16 calibration.

        \begin{table*}
          \centering
          \caption{Summary of the So12 and Te16 calibrations.}
         \label{tab:cal_comparison}
         \begin{adjustbox}{max width=\textwidth}
         \begin{tabular}{lccccc}
         \hline
         \hline
         Calibration &  ratios & \ion{Fe}{i}\_lines & \teff  & $\log g$  & [Fe/H]  \\ 
          &   &  & (K) & ($dex$) & ($dex$) \\ 
         \hline
         So12    &  $433$    & $151$ & [$4483, 6403$] & [$3.63, 4.92$] & [$-0.84, 0.39$] \\ 
         Te16       &  $322$    & $100$ & [$4483, 6403$] & [$2.37, 4.92$] & [$-0.84, 0.39$] \\
         
         \hline
         \end{tabular}
         \end{adjustbox}
         \end{table*}

         \begin{table*}
          \centering
          \caption{Comparison of the So12 and Te16 calibrations and the spectroscopic values for the calibration samples.}
         \label{tab:full_comparison}
         \begin{adjustbox}{max width=\textwidth}
         \begin{tabular}{lccccccc}
         \hline
         \hline
         Calibration & Sample & $\Delta$\teff  & $\Delta$\teff\_median   & $\Delta$\teff\_$\sigma$  & $\Delta$[Fe/H]   & $\Delta$[Fe/H]\_median   & $\Delta$[Fe/H]\_$\sigma$   \\ 
           &   &  (K) &   (K) &  (K) &  ($dex$) &  ($dex$) &  ($dex$) \\ 
         \hline
         So12  &  So08   &   $-22$   &   $-21$   &    $69$    &      $0.00$      &      $0.00$       &      $0.04$     \\ 
         So12  &  Al15   &   $-208$  &   $-219$  &    $69$    &      $-0.29$     &      $-0.29$      &      $0.07$     \\
         So12  &  Joint  &   $-90$   &   $-61$   &    $112$   &      $-0.10$     &      $-0.02$      &      $0.15$     \\
         Te16  &  So08   &   $17$    &   $23$    &    $76$    &      $0.01$      &      $0.02$       &      $0.04$     \\
         Te16  &  Al15   &   $-32$   &   $-36$   &    $58$    &      $-0.07$     &      $-0.06$      &      $0.08$     \\
         Te16  &  Joint  &   $-1$    &   $-8$    &    $74$    &      $-0.02$     &      $0.00$       &      $0.07$     \\
         
         \hline
         \end{tabular}
         \end{adjustbox}
         \end{table*}

         \begin{table}
         \centering
         \caption{Limits of applicability of the Te16 calibration.}
         \label{tab:limits}
         
         \begin{tabular}{lcc}
         \hline
         \hline
         \teff & $\log g $  & [Fe/H]  \\ 
         (K) &  ($dex$) &  ($dex$) \\ 
         \hline
         $[ 4500 , 6500 ]$ & $[ 2.5 , 4.9 ]$ & $[ -0.8 , 0.5 ]$\\ 
         
         \hline
         \end{tabular}
         \end{table}

         \begin{table*}
          \centering
          \caption{Application of the Te16 calibration to the validation samples for stars within the applicability limits.}
         \label{tab:application_comparison}
         \begin{adjustbox}{max width=\textwidth}
         \begin{tabular}{lcccccc}
         \hline
         \hline
         Sample & $\Delta$\teff  & $\Delta$\teff\_median & $\Delta$\teff\_$\sigma$ & $\Delta$[Fe/H]  & $\Delta$[Fe/H]\_median  & $\Delta$[Fe/H]\_$\sigma$  \\ 
           &   (K) &   (K) &  (K) & ($dex$) & ($dex$) & ($dex$) \\ 
         \hline
         GES    &   $20$    &   $38$    &    $141$   &      $0.05$      &      $0.06$       &      $0.11$     \\ 
         Sa09   &   $-44$   &   $-41$   &    $75$    &      $-0.04$     &      $-0.06$      &      $0.10$     \\
         Gaia   &   $43$    &   $40$    &    $91$    &      $0.08$      &      $0.09$       &      $0.18$     \\
         So11    &   $47$    &   $46$    &    $89$    &      $0.02$      &      $0.02$       &      $0.05$     \\
         Combined-validation    &   $40$    &   $37$    &    $93$    &      $0.02$      &      $0.02$       &      $0.07$     \\
         
         \hline
         \end{tabular}
         \end{adjustbox}
         \end{table*}

        A final remark should be made concerning the dependence of our values on $\log g$. 
        Figures \ref{fig:teff_orig_joint} and \ref{fig:feh_orig_joint} are also colour-coded
        with $\log g$. 
        A trend for low-$\log g$ stars in the Te16 calibration (middle panel)  
        to have underestimated
        \teff and
        [Fe/H] (e.g. giants with $\log g < 2.5 \ dex$) is evident. Although
        this trend has been greatly minimized, it is still present and should be taken
        into account when we consider the results of the new calibration.
        A partial justification for these discrepancies 
        may be the fact that cooler stars in general 
        have higher uncertainties in the EW determination.
        It should also be pointed out that this underestimation is within the 
        standard deviation of the calibration.
        The limits of the Te16 calibration reflect the parameters of the calibration sample
        and are presented in
        Table \ref{tab:limits}.

        \subsection{Validation with independent samples}\label{subs:indsamp}
        
        After obtaining the Te16 calibration for \teff and [Fe/H], we applied it
        to four completely independent samples: the Sa09, Gaia, GES, and So11 samples.

        Figure \ref{fig:teff_testing_all} shows the results of \teff computed
        in this work against the \teff determined by spectroscopy for the four validation samples and is colour-coded
        to the $\log g$ value. Stars outside the limits of the Te16 calibration ($\log g < 2.5$)
        are not plotted in this figure.

        \begin{figure}
        
         \includegraphics[width=\columnwidth]{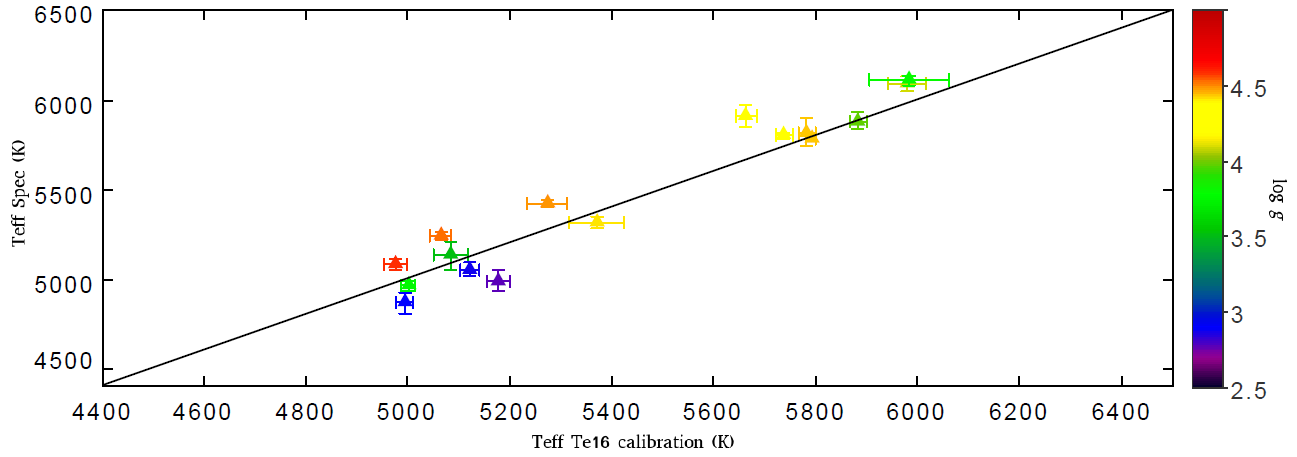}        
         \includegraphics[width=\columnwidth]{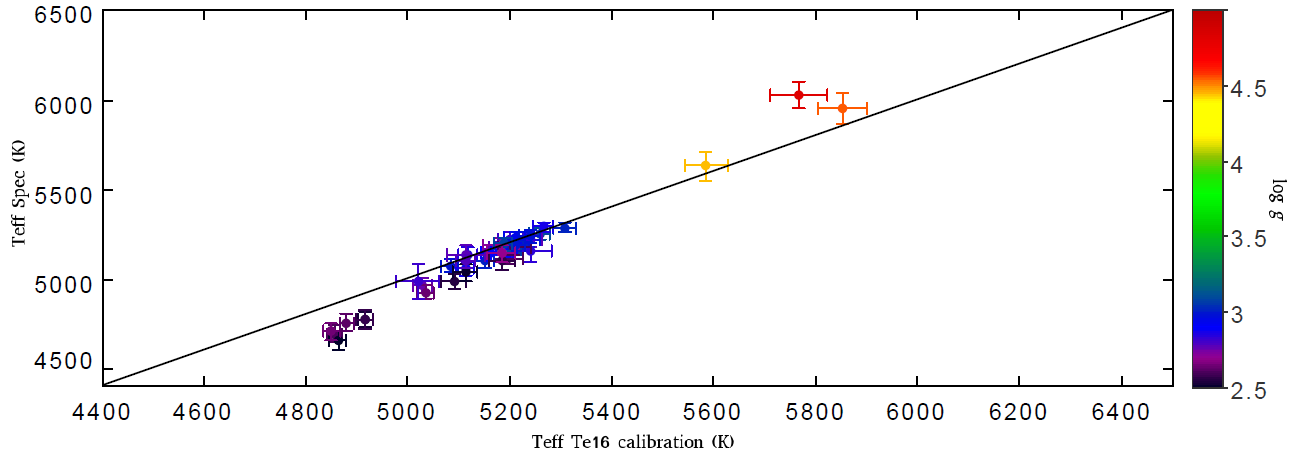}        
         \includegraphics[width=\columnwidth]{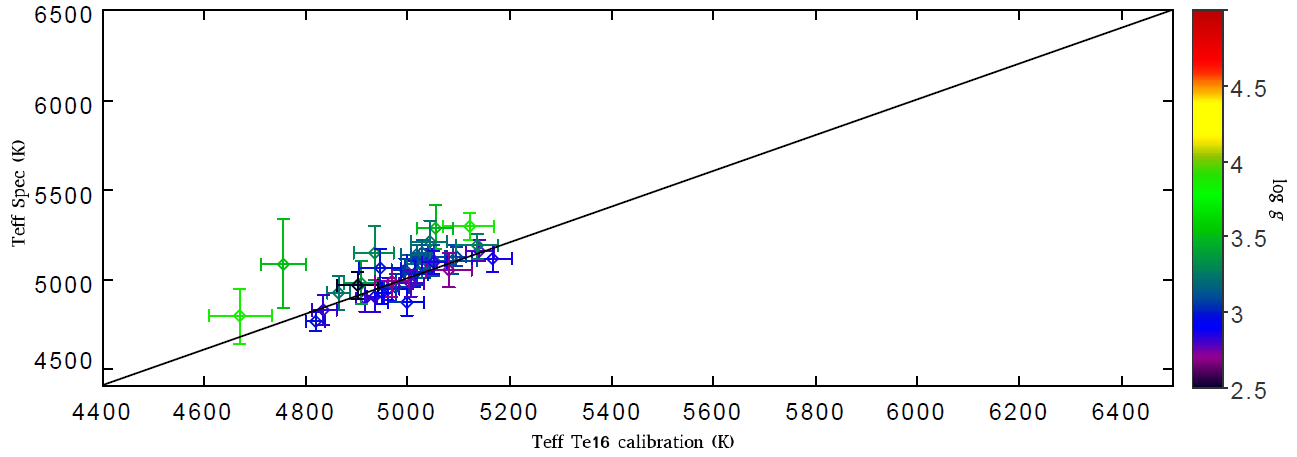} 
         \includegraphics[width=\columnwidth]{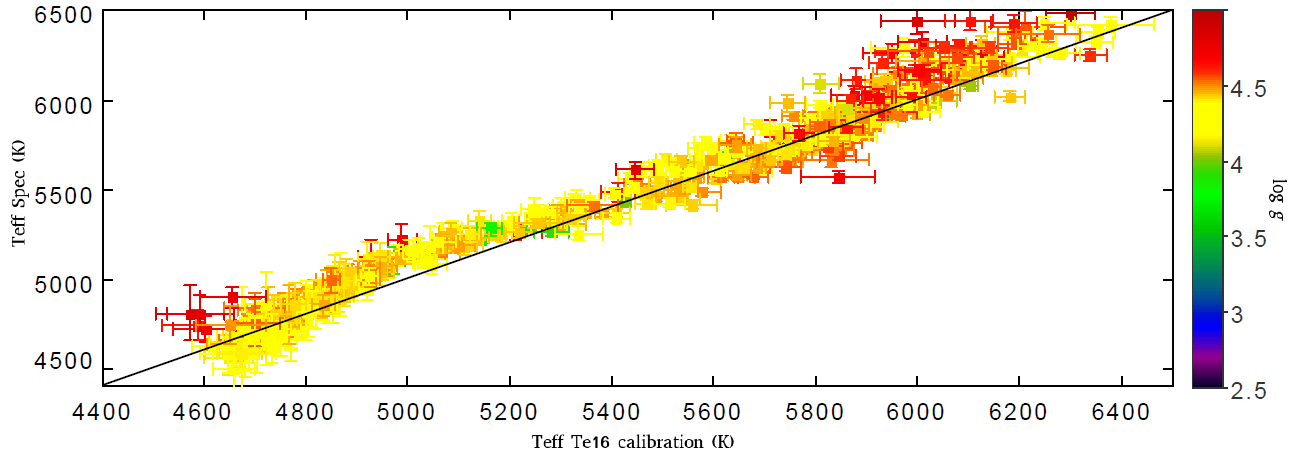}                 
         \caption{Comparison between the \teff computed with TMCalc and the spectroscopic
         values with the Te16 calibration for the
         Gaia, Sa09, GES, and So11 sample from top to
bottom. 
         The colour-code represents the $\log g$ values and the black
         line represents the identity line.
         Only stars within the limits of applicability are plotted.}
        \label{fig:teff_testing_all}
        \end{figure}

        The results for the [Fe/H] computation of the independent samples
        are shown in Fig. \ref{fig:feh_testing_all},
        the same goodness-of-fit is evident in this plot for both samples.
        Again we encounter some outliers, but it should be made clear that these are stars with
        low-$\log g$ values and \teff and, therefore, close to our applicability
        limits (see Table \ref{tab:limits}).

        \begin{figure}
        
         \includegraphics[width=\columnwidth]{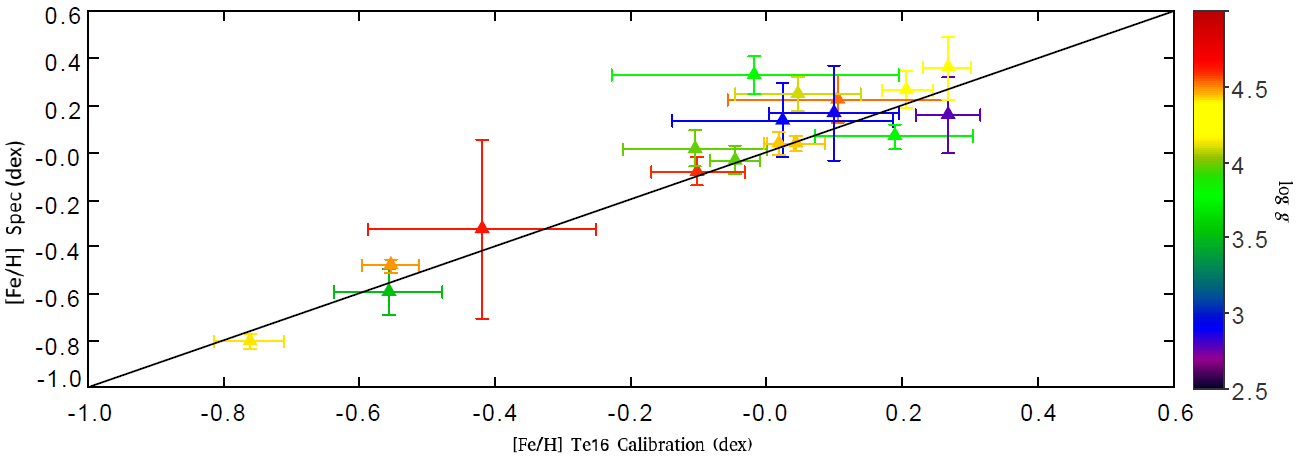} 
         \includegraphics[width=\columnwidth]{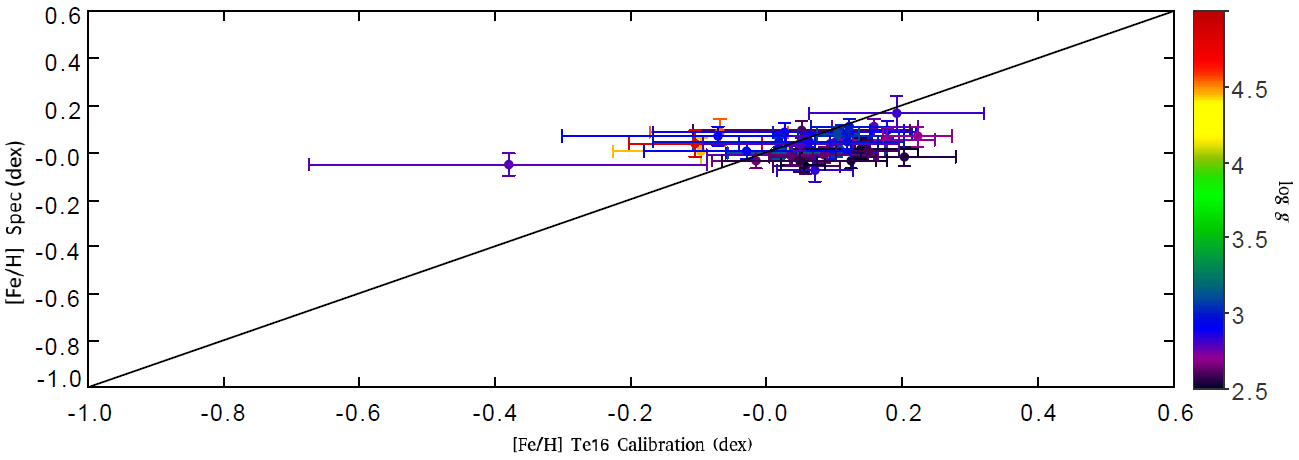} 
         \includegraphics[width=\columnwidth]{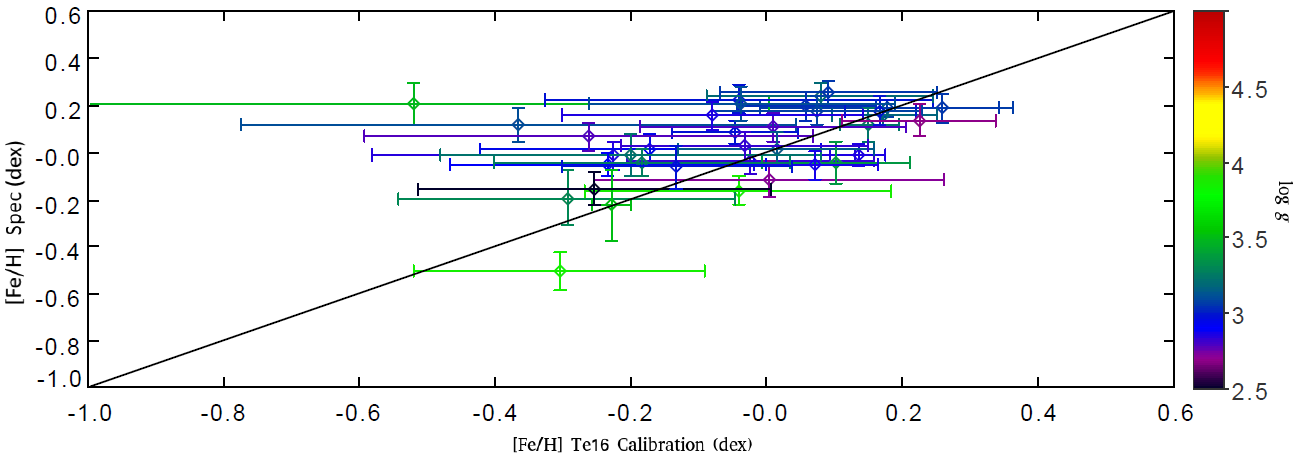}  
         \includegraphics[width=\columnwidth]{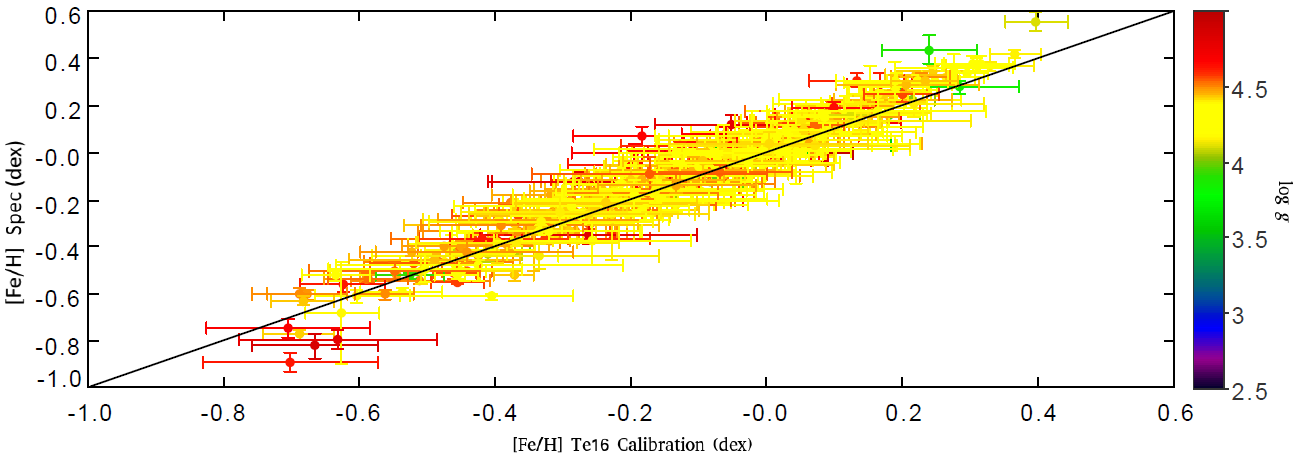}          
         \caption{Comparison between the [Fe/H] computed with TMCalc and the spectroscopic
         values with the Te16 calibration for       Gaia, Sa09, GES, and So11 sample. 
         The colour-code represents the $\log g$ values and the black
         line represents the identity line.
         Only stars within the limits of applicability are plotted.}
        \label{fig:feh_testing_all}
        \end{figure}

        A summary of the application of the Te16 calibration to the various validation samples
        is provided in Table \ref{tab:application_comparison}.

        Applying the Te16 calibration to stars outside the applicability limits
        will introduce higher uncertainties.    
        These applicability limits raise
        the question of how one should proceed when there is no prior knowledge
        of the $\log g$ of a star. Figure \ref{fig:errors_ind} shows the errors computed
        by TMCalc for \teff and [Fe/H] as a function of $\log g$ for the Gaia, Sa09, and joint samples
        before the selection based on the applicability limits. 
        Based on an empirical 
        analysis of these plots, we propose that stars that simultaneously have [Fe/H] errors greater than
        $0.1 \ dex$ and \teff errors greater than $18 \ K$ should be flagged for further examination because they may be outside, or close to, our applicability limits.
        The high errors can also be mimicked by low (S/N) spectra, therefore, this criterion 
        is only an indication that the flagged stars need to
be more carefully analysed.
        We tested this criterion and found that we would flag the stars with $\log g < 2.5$
        and obtain 62 stars ($\sim 10\%$), 4 stars ($\sim 19\%$), and 12 stars ($\sim 22\%$) as false positives, that is, stars that were flagged
        as suspicious but are not outside our applicability limits for the joint, Gaia, and Sa09 samples, respectively.
        
        The errors obtained by TMCalc are computed considering the dispersion 
        of the values given by all the individual calibrations, as independent of all others 
        and, therefore, they are divided by the square root of the number of individual calibrations used.
        The error in [Fe/H] is obtained by using the 1-$\sigma$ temperature error.
        The final error is obtained from the quadratic sum of the two error sources \citep{sousaTMCALC}.

        \begin{figure}
        
         \includegraphics[width=\columnwidth]{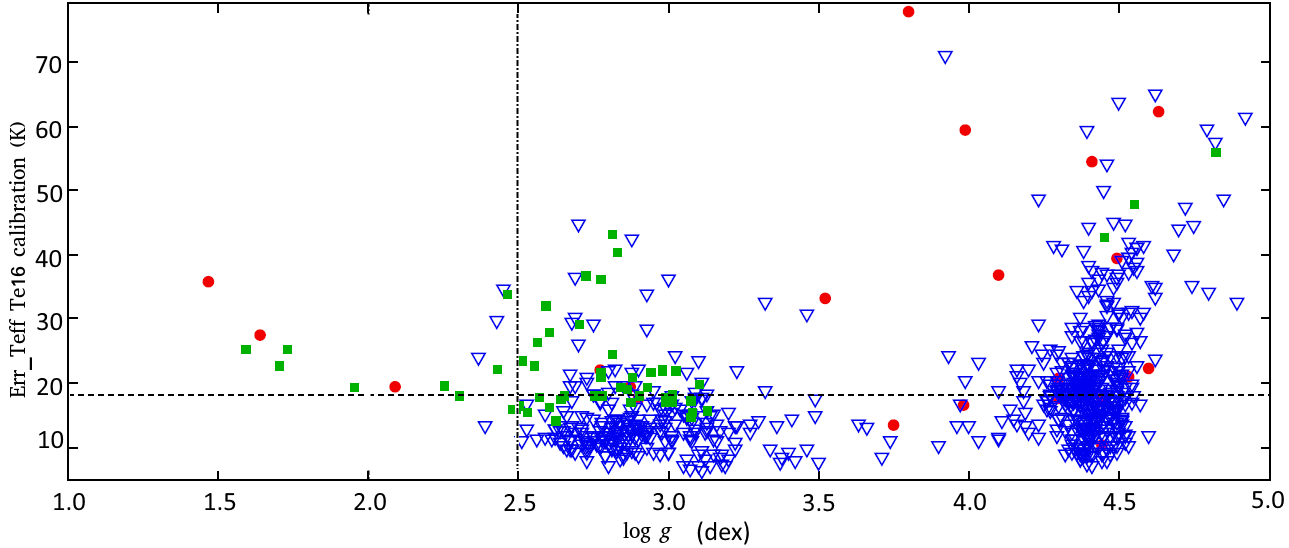}
         \includegraphics[width=\columnwidth]{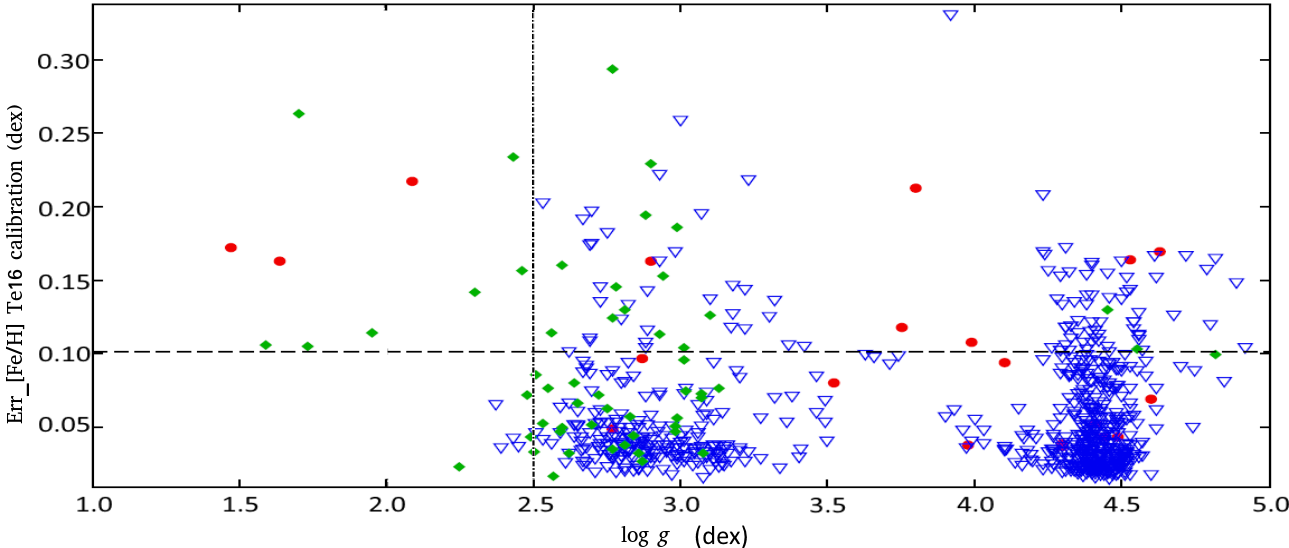}
         \caption{Errors in \teff(top panel) and [Fe/H] (bottom panel) as a function 
         of $\log g$. We plot the Gaia sample (red circles), the Sa09 sample (green diamonds),
         and the joint-sample (blue open triangles). The dashed lines represent the maximum error a measurement can have to
         be considered trustworthy, 18 K and 0.10 dex for \teff and [Fe/H]. The vertical
         dot-dashed lines represent the $\log g $ limit of the Te16 calibration.}
        \label{fig:errors_ind}
        \end{figure}    
        
        \subsection{Effect of \teff errors on [Fe/H]}\label{subs:impact}
        
        As we discussed in Sect. \ref{ch:metalcal},
        the [Fe/H] calibration has a non-negligible dependence on \teff.
TMCalc first computes the \teff of a given star and then applies
        that value in the determination of [Fe/H].
        To understand the effect of a poorly determined \teff 
        on the computations of [Fe/H], we performed the following test on the joint sample:
        we computed the [Fe/H] using the ARES+MOOG \teff instead of the \teff
        obtained from the application of the Te16 calibration.
        
        Figure \ref{fig:feh_impact} shows the result of this computation.
        The [Fe/H] differences are below the 2-$\sigma$ of
        the new calibration and are therefore not significantly 
        affected by the \teff determination.

        Additionally, Fig. \ref{fig:feh_impact} is also colour-coded to show the spectroscopic
        temperatures. With the exception of a few cooler stars, the dependence between
        \teff and [Fe/H] computed appears to have a linear behaviour. When we consider
        that the [Fe/H] is computed by inverting Eq. \ref{eq:metal}, the apparent linearity
        is not a surprise because the inverted equation has a  first-order dependence on \teff.
        
        \begin{figure}
        
         \includegraphics[width=\columnwidth]{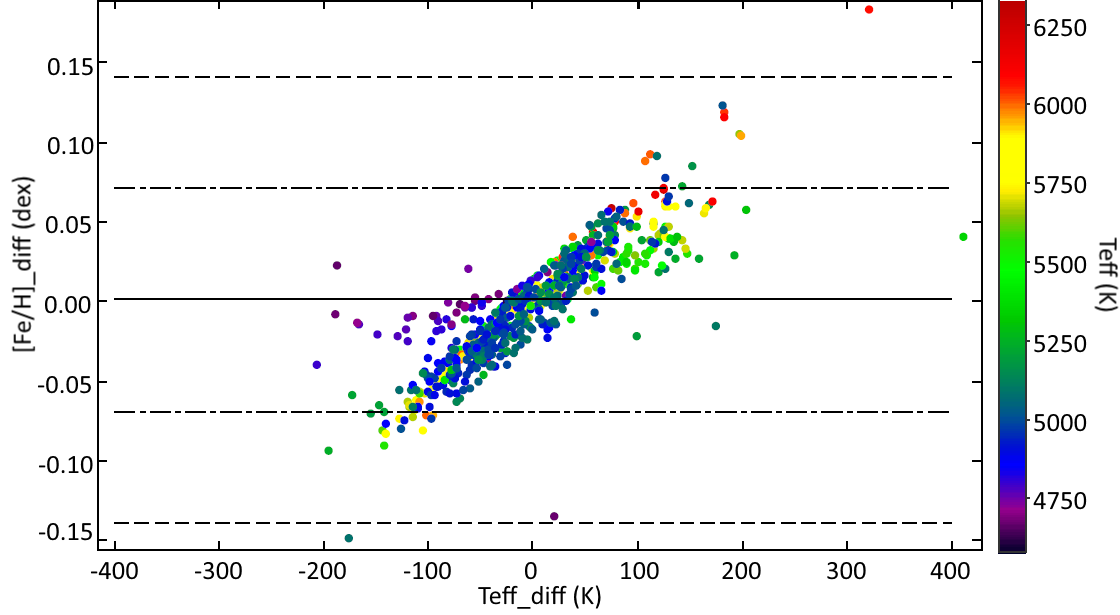}
         \caption{Effect of the spectroscopic \teff on the [Fe/H] determination. 
         The horizontal axis shows the difference
         between the  spectroscopic value of \teff and the one computed by TMCalc. 
         The vertical axis shows the difference between the original computation
         of [Fe/H] and the value computed with the spectroscopic \teff.
         The black line represents the zero of the vertical axis, the semi-dashed and the 
         dashed black lines
         represent the 1-$\sigma$ and 2-$\sigma$ of the Te16 calibration, respectively.}
        \label{fig:feh_impact}
        \end{figure}

\section{Summary and conclusions}\label{ch:conclusions}
        
        We presented new calibrations to obtain \teff and [Fe/H] from
        the EWs of stellar spectra. These new calibrations are the first to
        successfully accommodate both FGK dwarfs and GK giants simultaneously,
        covering an increased range of spectral types and evolutionary
        stages. 
        Our careful selection of
        the calibration sample and our improvement of the calibration method
        itself produced the calibrations presented in this work.

        Using a joint sample of 451 FGK dwarfs and 257 GK giants, we determined
        new calibrations. These were applied and successfully 
        tested against the pre-existing
        calibrations using the same sample. The Te16 calibration was applied to four different
        validation samples and was found to be effective within
        the range of $4500 K < \teff < 6500 K$, $2.5 < \log g <4.9$ dex, and $-0.8 < [Fe/H]<0.5$ dex,
        which covers most of FGK dwarfs and GK giants.

        We showed that the dependence of the value of [Fe/H] on \teff is within
        the error bars and therefore cannot be used to explain
the poor estimates of [Fe/H] in the validation sample. 
        This probably is the reason why these stars are outside our
        application range. We proposed a method to flag stars outside
        our applicability limits based on errors.
        
        Future work should be focused on increasing the 
        range of values of the calibration sample and, therefore,
        increasing the applicability limits of the calibration.
        Additional corrections are required to account
        for low-$\log g$ stars.
        
        We built a Python code, GeTCal, that is capable of obtaining \teff and [Fe/H] calibrations
        for any given sample of calibration stars. This program produces
        calibration files compatible with the existing TMCalc code and will therefore
        be distributed with future versions of TMCalc.

        This work provides a fast way to determine stellar 
        atmospheric parameters from spectrographic observations of FGK-dwarf and GK-giant 
        stars. 
        This calibration can be used to select targets for observations with future spectrographs such as ESPRESSO \citep{espress14}. \
        It can also be expanded to other spectral types by 
        applying the GeTCal code and an appropriate calibration sample. 
        
\begin{acknowledgements}
G.D.C.T. was supported by a research fellowship Ref: CAUP2013-05UnI-BI,
funded by the European Commission through project SPACEINN (FP7-SPACE-2012-312844)
and by an FCT/Portugal PhD grant PD/BD/113478/2015.
M.J.M. was supported in part by FCT/Portugal through projects PEst-C/FIS/UI0003/2013
and UID/FIS/04434/2013.
N.C.S. was supported by FCT through the Investigador FCT contract reference 
IF/00169/2012 and POPH/FSE (EC) by FEDER funding through the program Programa
Operacional de Factores de Competitividade - COMPETE.
S.G.S acknowledges the support from the Funda\c{c}\~ao para a Ci\^encia e Tecnologia,
FCT (Portugal) and POPH/FSE (EC), in the form of the Investigador FCT contract
reference IF/00028/2014.
G.I. acknowledges financial support from the Spanish Ministry project MINECO AYA2011-29060.
This work is supported by the European Research Council/European Community under the FP7
through Starting Grant agreement number 239953.
\end{acknowledgements}

\bibliographystyle{aa}

\begin{thebibliography}{}

\bibitem[Alves et al.(2015)]{alves14} Alves, S., Benamati, L., Santos, N.~C., et al.\ 2015, \mnras, 448, 2749 

\bibitem[Bensby et al.(2014)]{bens14} Bensby, T., Feltzing, S., Oey, M.~S.,\ 2014, \aap, 562, A71

\bibitem[Bland-Hawthorn et al.(2014)]{gala} Bland-Hawthorn, J., Sharma, S., \& Freeman, K.\ 2014, EAS Publications Series, 67, 219

\bibitem[Casagrande et al.(2010)]{casa10} Casagrande, L., Ram{\'{\i}}rez, I., Mel{\'e}ndez, J., Bessell, M., Asplund, M.,\ 2010, \aap, 512, A54 

\bibitem[Chaplin et al.(2014)]{chap14} Chaplin, W.~J. et al.,\ 2014, \apjs , 210, 1 

\bibitem[Edvardsson et al.(1993)]{edv93} Edvardsson, B., Andersen, J., Gustafsson, B., Lambert, D.~L., Nissen, P.~E., Tomkin, J.,\ 1993, \aap, 275, 101 

\bibitem[Fischer et al.(1993)]{fisch05} Fischer, D.~A., Valenti, J.,\ 2005, \apj, 622, 1102 

\bibitem[Gilmore et al.(2012)]{ges} Gilmore, G., Randich, S., Asplund, M., et al.\ 2012, The Messenger, 147, 25 

\bibitem[Gray(1996)]{gray96} Gray, D.~F.\ 1996, Stellar Surface Structure, 176, 227 

\bibitem[Gray (2005)]{graybook} Gray, D.~F.,\ 2005, The Observation and Analysis of Stellar Photospheres, 3rd Edition, ~ISBN 0521851866.~, UK: Cambridge University Press, 2005

\bibitem[Heiter et al.(2015)]{heit15} Heiter, U., Jofr{\'e}, P., Gustafsson, B., et al.\ 2015, \aap, 582, A49 

\bibitem[Huber et al.(2012)]{hub12} Huber, D. et al.,\ 2012, \apj, 760, 32

\bibitem[Jofr{\'e} et al.(2014)]{jofre14} Jofr{\'e}, P., Heiter, U., Soubiran, C., Blanco-Cuaresma, S. et al., \ 2014, \aap, 564, A133 

\bibitem[Kovtyukh et al.(2003)]{kov03} Kovtyukh , V.~V., Soubiran, C., Belik, S.~I., Gorlova, N.~I.,\ 2003, \aap, 411, 559 

\bibitem[Mortier et al.(2013)]{mortier13} Mortier, A., Santos, N.~C., Sousa, S.~G., Adibekyan, V.~Z., Delgado Mena, E., Tsantaki, M., Israelian, G., Mayor, M.,\ 2013, \aap, 557, A70 

\bibitem[Pepe et al.(2014)]{espress14} Pepe, F., Molaro, P., Cristiani, S., et al.\ 2014, Astronomische Nachrichten, 335, 8 

\bibitem[Santos et al.(2009)]{santos09} Santos, N.~C., Lovis, C., Pace, G., Melendez, J., Naef, D.,\ 2009, \aap, 493, 309 

\bibitem[Santos et al.(2012)]{santos12} Santos, N.~C., Lovis, C., Melendez, J. , Montalto, M., Naef, D., Pace, G.,\ 2012, \aap, 538, A151

\bibitem[Sousa et al.(2007)]{sousaARES} Sousa, S.~G., Santos, N.~C., Israelian, G., Mayor, M., \& Monteiro, M.~J.~P.~F.~G.,\ 2007, \aap, 469, 783 

\bibitem[Sousa et al.(2008)]{sousa451stars} Sousa, S.~G., Santos, N.~C., Mayor, M., et al.,\ 2008, \aap, 487, 373 

\bibitem[Sousa et al.(2010)]{sousa10} Sousa, S.~G., Alapini, A., Israelian, G., \& Santos, N.~C.,\ 2010, \aap, 512, A13 

\bibitem[Sousa et al.(2011)]{sousa11} Sousa, S.~G., Santos, N.~C., Israelian, G., Mayor, M., \& Udry, S.\ 2011, \aap, 533, A141 

\bibitem[Sousa et al.(2012)]{sousaTMCALC} Sousa, S.~G., Santos, N.~C., \& Israelian, G.,\ 2012, \aap, 544, A122 

\bibitem[Sousa(2014)]{sousaschool} Sousa, S., 2014,  arXiv:1407.5817 

\bibitem[Sousa et al.(2015)]{ares2015} Sousa, S.~G., Santos, N.~C., Adibekyan, V., Delgado-Mena, E., \& Israelian, G.\ 2015, \aap, 577, A67 

\bibitem[Sneden (1973)]{sneden73moog} Sneden, C.~A.\ 1973, Ph.D. Thesis, The University of Texas at Austin

\bibitem[Steinmetz et al.(2006)]{rave} Steinmetz, M., Zwitter, T., Siebert, A., et al.\ 2006, \aj, 132, 1645

\bibitem[Tsantaki et al.(2013)]{maria13} Tsantaki, M., Sousa, S.~G., Adibekyan, V.~Z., et al.,\ 2013, \aap, 555, A150 

\bibitem[Tsantaki et al.(2014)]{maria14} Tsantaki, M., Sousa, S.~G., Santos, N.~C., Montalto, M., Delgado-Mena, E., Mortier, A., Adibekyan, V.~Z., Israelian, G., \ 2014, \aap, 570, A80

\bibitem[Upton et al.(1996)]{upton96}  Upton, G., Cook, I.,\ 1996, Understanding Statistics. Oxford University Press.

\end{thebibliography}

\end{document}